\documentclass[conference]{IEEEtran}
\usepackage{ifpdf}
\usepackage[nospace]{cite}
\usepackage{graphicx}
\usepackage{bm}
\usepackage[cmex10]{amsmath}
\usepackage{amssymb}
\usepackage{array}
\usepackage{syntonly}
\usepackage{amsthm}
\usepackage{amsfonts}
\usepackage{epsfig}
\usepackage{latexsym}

\usepackage{fixltx2e}
\usepackage{url}
\theoremstyle{definition} 
\theoremstyle{definition} 
\theoremstyle{definition} 
\begin{document}
\title{{\LARGE Power Allocation in MIMO Wiretap Channel with Statistical CSI 
and Finite-Alphabet Input} }
\author{\IEEEauthorblockN{Sanjay Vishwakarma}
\IEEEauthorblockA{Dept. of ECE\\
Indian Institute of Science\\
Bangalore 560012 \\
Email: sanjay@ece.iisc.ernet.in}
\and
\IEEEauthorblockN{A. Chockalingam}
\IEEEauthorblockA{Dept. of ECE\\
Indian Institute of Science\\
Bangalore 560012 \\
Email: achockal@ece.iisc.ernet.in}}
\vspace{-30mm}
\author{{\large Sanjay Vishwakarma and A. Chockalingam} \\
{\normalsize Department of ECE, Indian Institute of Science,
Bangalore 560012 }
}
\maketitle
\begin{abstract}
In this paper, we consider the problem of power allocation in MIMO wiretap 
channel for secrecy in the presence of multiple eavesdroppers. Perfect 
knowledge of the destination channel state information (CSI) and only the 
statistical knowledge of the eavesdroppers CSI are assumed. We first consider 
the MIMO wiretap channel with Gaussian input. Using Jensen's inequality, we 
transform the secrecy rate max-min optimization problem to a single maximization 
problem. We use generalized singular value decomposition and transform the 
problem to a concave maximization problem which maximizes the sum secrecy rate 
of scalar wiretap channels subject to linear constraints on the transmit 
covariance matrix. We then consider the MIMO wiretap channel with 
finite-alphabet input. We show that the transmit covariance matrix obtained 
for the case of Gaussian input, when used in the MIMO wiretap channel with 
finite-alphabet input, can lead to zero secrecy rate at high transmit powers. 
We then propose a power allocation scheme with an additional power constraint 
which alleviates this secrecy rate loss problem, and gives non-zero secrecy 
rates at high transmit powers. 
\end{abstract}
{\em keywords:}
{\em {\footnotesize
MIMO wiretap channel, physical layer security, secrecy rate, multiple 
eavesdroppers, statistical CSI, finite-alphabet input.
}} 
\IEEEpeerreviewmaketitle
\section{Introduction}
\label{sec1}
Wireless transmissions are vulnerable to eavesdropping due to their broadcast 
nature. There is a growing demand to address the issue of providing security 
in wireless networks. Secrecy in wireless communication networks can be 
achieved using physical layer techniques, where the legitimate receiver gets 
the transmitted information correctly and the eavesdropper receives no or very 
little information. Achievable secrecy rates and secrecy capacity bounds for 
multiple antenna point-to-point wiretap channel has been studied in 
\cite{ic1,ic2,ic3,ic4}. In \cite{ic1}, \cite{ic5}, multiple-input single-output
(MISO) wiretap channel is considered, and secrecy rate is computed assuming 
statistical information of the eavesdropper channel. In \cite{ic6}, \cite{ic7}, 
secrecy capacity of the multiple-input multiple-output (MIMO) wiretap channel 
has been computed assuming perfect channel state information (CSI) knowledge 
of the destination and the eavesdropper. These works consider secrecy rate 
when the input to the channel is Gaussian. In practice, the input to the 
channel will be from a finite alphabet set, e.g., $M$-ary alphabets.
The effect of finite-alphabet input on the achievable secrecy rate for
various channels has been studied in \cite{ic8,ic9,ic10,ic11}. It has been 
shown that with finite-alphabet input, increasing the power beyond a maximum 
point is harmful as the secrecy rate curve dips continuously thereafter.
In \cite{ic12}, design of optimum linear transmit precoding for maximum 
secrecy rate over MIMO wiretap channel with finite-alphabet input and with 
perfect eavesdropper CSI assumption has been investigated. 

In this paper, we consider the problem of power allocation in MIMO wiretap channel
for secrecy in the presence of multiple eavesdroppers. To our knowledge, such 
a study for the case of finite-alphabet input when only the statistical CSI of 
the eavesdroppers is assumed has not been reported before. Our approach to
study this problem, which is adopted in this paper, is summarized as follows. 
First, we consider the MIMO wiretap channel with Gaussian input and knowledge
of statistical CSI of the eavesdroppers. We transform the secrecy rate max-min 
optimization problem with Gaussian input into a single maximization problem 
using Jensen's inequality. Generalized singular value decomposition (GSVD) is 
used to transform the problem to a concave maximization problem which maximizes 
the sum secrecy rate of scalar wiretap channels subject to linear constraints 
on the transmit covariance matrix. We then consider the MIMO wiretap channel 
with finite-alphabet input and knowledge of statistical CSI of the eavesdroppers. 
It is found that when the transmit covariance matrix obtained for the case of 
Gaussian input is used in the MIMO wiretap channel with finite-alphabet input, the 
secrecy rate goes to zero at high transmit powers. Therefore, we propose a 
power allocation scheme with an additional power constraint to deal with
this secrecy rate loss. The proposed scheme is shown to alleviate the secrecy 
rate loss problem and gives non-zero secrecy rates at high transmit powers.

The rest of the paper is organized as follows. The system model is presented
in Section \ref{sec2}. The secrecy rate with Gaussian input is studied in
Section \ref{sec3}. The secrecy rate with finite-alphabet input is studied in
Section \ref{sec4}. Numerical results and conclusions are presented in 
Section \ref{sec5} and Section \ref{sec6}, respectively. 

$\bf{Notations:}$ Vectors are denoted by boldface lower case letters, and 
matrices are denoted by boldface upper case letters. $\boldsymbol{A} \in 
\mathbb{C}^{N_{1} \times N_{2}}$ implies that $\boldsymbol{A}$ is a complex 
matrix of dimension $N_{1} \times N_{2}$. $\boldsymbol{A} \succeq \boldsymbol{0}$ 
denotes that $\boldsymbol{A}$ is a positive semidefinite matrix.
$\boldsymbol{I}$ denotes the identity matrix. 
Transpose and complex conjugate transpose operations are denoted with 
$[.]^{T}$ and $[.]^{\ast}$, respectively. $diag(\boldsymbol{a})$ denotes a 
diagonal matrix with elements of vector $\boldsymbol{a}$ on the diagonal of 
the matrix. $diag(\boldsymbol{A})$ denotes a vector formed with the diagonal
entries of matrix $\boldsymbol{A}$.
$\mathbb{E}\big[.\big]$ denotes expectation operation. 

\section{System Model}
\label{sec2}
\begin{figure}
\center
\includegraphics[totalheight=1.8in,width=3.20in]{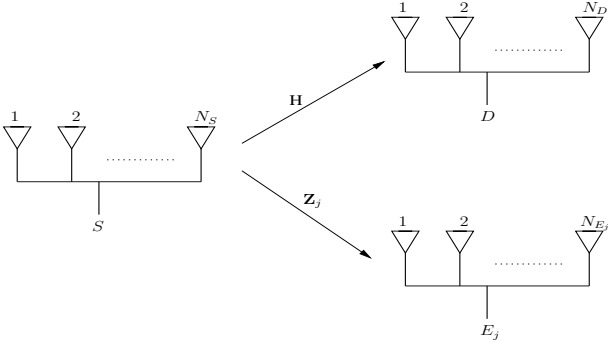}
\caption{System model.}
\label{fig1}
\vspace{-4mm}
\end{figure}

Consider a MIMO wiretap channel which consists of a source $S$, an intended 
destination $D$, and $J$ eavesdroppers \{$E^{}_{1},E^{}_{2},\cdots,E^{}_{J_{}}$\}.
The system model is shown in Fig. \ref{fig1}. Source $S$ has $N_{S}$ transmit 
antennas, destination $D$ has $N_{D}$ receive antennas and each eavesdropper 
$E_{j}$ has $N_{E_{j}}$ receive antennas. The complex fading channel gain 
matrix between $S$ to $D$ is denoted by 
$\boldsymbol{H} \in \mathbb{C}^{N_{D} \times N_{S}}$. Likewise, the channel 
gain matrix between $S$ to $E_{j}$ is denoted by 
$\boldsymbol{Z}_{j} \in \mathbb{C}^{N_{E_{j}} \times N_{S}}$. We assume that 
the channel gain matrix, $\boldsymbol{H}$, between $S$ to $D$ is known 
perfectly. We also assume that the channel gains of all the eavesdroppers 
are unknown and that all the channel gains of eavesdroppers are i.i.d 
$\sim \mathcal{CN}(0, \sigma^{2}_{E_{j}})$.
Let $P_{0}$ denote the total available transmit power. The source $S$ transmits 
the complex vector symbol $\boldsymbol{x} \sim \mathcal{CN}(\boldsymbol{0}, \boldsymbol{Q})$,
where $\boldsymbol{Q} = \mathbb{E}\{ \boldsymbol{x} \boldsymbol{x}^{\ast} \}$ is 
the transmit covariance matrix and $trace(\boldsymbol{Q}) \leq P_{0}$. Let 
$\boldsymbol{y}_{D_{}}$ and $\boldsymbol{y}_{E_{j}}$ denote the 
received signals at the destination $D$ and the $j$th eavesdropper
$E^{}_{j}$, respectively. We then have
\begin{eqnarray}
\boldsymbol{y}_{D}=\boldsymbol{H}\boldsymbol{x}+\boldsymbol{\eta}_{D}, \label{eqn1} \\
\boldsymbol{y}_{E_{j}}=\boldsymbol{Z}_{j}\boldsymbol{x}+\boldsymbol{\eta}_{E_{j}}, 
\label{eqn2}
\end{eqnarray}
where $\boldsymbol{\eta}_{D} \sim \mathcal{CN}(\boldsymbol{0}, N_{0}\boldsymbol{I})$ and 
$\boldsymbol{\eta}_{E_{j}} \sim \mathcal{CN}(\boldsymbol{0}, N_{0}\boldsymbol{I})$ are 
the i.i.d. noise vectors at $D$ and $E_{j}$, respectively. 

\section{MIMO Wiretap Channel with Gaussian Input}
\label{sec3}
For a given $\boldsymbol{H}$, using (\ref{eqn1}), the information rate at 
the destination $D$ is
\begin{eqnarray}
I(\boldsymbol{x}; \ \boldsymbol{y}_{D}) = \log_{2} \det\Big(\boldsymbol{I} + \frac{\boldsymbol{H}\boldsymbol{Q}\boldsymbol{H}^{\ast}}{N_{0}}\Big). \label{eqn3}
\end{eqnarray}
Similarly, for a given $\boldsymbol{Z}_{j}$, using (\ref{eqn2}), the information 
rate at the $j$th eavesdropper $E_{j}$, $\forall j = 1,2,\cdots,J,$ is
\begin{eqnarray}
I(\boldsymbol{x}; \ \boldsymbol{y}_{E_{j}}) = \log_{2} \det\Big(\boldsymbol{I} + \frac{\boldsymbol{Z}_{j}\boldsymbol{Q}\boldsymbol{Z}^{\ast}_{j}}{N_{0}}\Big). 
\label{eqn4}
\end{eqnarray}
Subject to the total power constraint $P_{0}$, using (\ref{eqn3}) and 
(\ref{eqn4}), the secrecy rate $R^{}_{s}$ for the MIMO wiretap channel is 
obtained by solving the following optimization problem \cite{ic1}, \cite{ic5}:
\begin{eqnarray}
R^{}_{s} \ = \ \max_{\boldsymbol{Q}} \ \min_{j : 1,2,\cdots,J} \bigg \{ \log_{2} \det\Big(\boldsymbol{I} + \frac{\boldsymbol{H}\boldsymbol{Q}\boldsymbol{H}^{\ast}}{N_{0}}\Big) - \nonumber \\ \mathbb{E}\bigg[\log_{2} \det\Big(\boldsymbol{I} + \frac{\boldsymbol{Z}_{j}\boldsymbol{Q}\boldsymbol{Z}^{\ast}_{j}}{N_{0}}\Big) \bigg] \bigg\}, \label{eqn5} \\
\geq \ \max_{\boldsymbol{Q}} \ \min_{j : 1,2,\cdots,J} \ \bigg\{\log_{2} \det\Big(\boldsymbol{I} + \frac{\boldsymbol{H}\boldsymbol{Q}\boldsymbol{H}^{\ast}}{N_{0}}\Big) - \nonumber \\ \log_{2} \det\Big(\boldsymbol{I} + \frac{ N_{E_{j}} \sigma^{2}_{E_{j}}\boldsymbol{Q}}{N_{0}}\Big) \bigg\}, \label{eqn12} \\
= \ \max_{\boldsymbol{Q}} \ \bigg\{\log_{2} \det\Big(\boldsymbol{I} + \frac{\boldsymbol{H}\boldsymbol{Q}\boldsymbol{H}^{\ast}}{N_{0}}\Big) - \nonumber \\ \log_{2} \det\Big(\boldsymbol{I} + \frac{ N_{E_{j_{0}}} \sigma^{2}_{E_{j_{0}}}\boldsymbol{Q}}{N_{0}}\Big)\bigg\}, \label{eqn13} \\
= \ \max_{\boldsymbol{Q}} \ \bigg\{\log_{2} \det\Big(\boldsymbol{I} + \frac{\boldsymbol{H}\boldsymbol{Q}\boldsymbol{H}^{\ast}}{N_{0}}\Big) - \nonumber \\ \log_{2} \det\Big(\boldsymbol{I} + \frac{ \boldsymbol{Z} \boldsymbol{Q}\boldsymbol{Z}^{\ast}}{N_{0}} \Big)\bigg\}, \label{eqn14}
\end{eqnarray}
\begin{eqnarray}
\text{s.t.} \quad \boldsymbol{Q} \ \succeq \ \boldsymbol{0}, \quad trace(\boldsymbol{Q}) \ \leq \ P_{0}, \label{eqn6}
\end{eqnarray}
where (\ref{eqn12}) is written using Jensen's inequality, 
$j_{0}$ in (\ref{eqn13}) corresponds to the eavesdropper with maximum 
$N_{E_{j_{0}}} \sigma^{2}_{E_{j_{0}}}$, and $\boldsymbol{Z}$ in 
(\ref{eqn14}) is $\sqrt{N_{E_{j_{0}}} \sigma^{2}_{E_{j_{0}}}} \boldsymbol{I}$. 
We intend to find the $\boldsymbol{Q}$ which maximizes the objective function 
in (\ref{eqn14}) subject to the constraints in (\ref{eqn6}). To do this, we 
take the GSVD \cite{ic13} of $\boldsymbol{H}$ and $\boldsymbol{Z}$ as 
\begin{eqnarray}
\boldsymbol{H} \ = \ \boldsymbol{U} \boldsymbol{\Lambda_{\boldsymbol{H}}}\Big[\boldsymbol{\Phi}^{\ast}\boldsymbol{T}, \ \boldsymbol{0}\Big]\boldsymbol{W}^{\ast} 
\label{eqn15} \\
\boldsymbol{Z} \ = \ \boldsymbol{V} \boldsymbol{\Lambda_{\boldsymbol{Z}}}\Big[\boldsymbol{\Phi}^{\ast}\boldsymbol{T}, \ \boldsymbol{0}\Big]\boldsymbol{W}^{\ast}.
\label{eqn16}
\end{eqnarray}
$\boldsymbol{U}$, $\boldsymbol{V}$, $\boldsymbol{\Phi}$, and $\boldsymbol{W}$ 
are unitary matrices of dimensions $N_{D} \times N_{D}$, $N_{S} \times N_{S}$, 
$k \times k$, and $N_{S} \times N_{S}$, respectively. 
$\boldsymbol{T}$ is an upper triangular matrix of size $k \times k$ and rank-$k$. 
$\boldsymbol{\Lambda_{\boldsymbol{H}}}$ and $\boldsymbol{\Lambda_{\boldsymbol{Z}}}$ 
are diagonal matrices of dimensions $N_{D} \times k$ and $N_{S} \times k$, 
respectively, and satisfy the condition
\begin{eqnarray}
\boldsymbol{\Lambda}^{T}_{\boldsymbol{H}} \boldsymbol{\Lambda}_{\boldsymbol{H}} + \boldsymbol{\Lambda}^{T}_{\boldsymbol{Z}} \boldsymbol{\Lambda}_{\boldsymbol{Z}} = \boldsymbol{I}.
\label{eqn16}
\end{eqnarray}
Substituting the GSVDs of $\boldsymbol{H}$ and $\boldsymbol{D}$ in 
$(\ref{eqn14})$, we write the problem as 

{\footnotesize
\begin{eqnarray}
\max_{\boldsymbol{Q}} \nonumber \\
\Big\{ \log_{2} \det\bigg(\boldsymbol{I} + \frac{\boldsymbol{U} \boldsymbol{\Lambda_{\boldsymbol{H}}}\Big[\boldsymbol{\Phi}^{\ast}\boldsymbol{T}, \ \boldsymbol{0}\Big]\boldsymbol{W}^{\ast}\boldsymbol{Q}\boldsymbol{W}^{}\Big[\boldsymbol{\Phi}^{\ast}\boldsymbol{T}, \ \boldsymbol{0}\Big]^{\ast}\boldsymbol{\Lambda}^{T}_{\boldsymbol{H}}\boldsymbol{U}^{\ast}}{N_{0}}\bigg)  \nonumber \\
- \log_{2} \det\bigg(\boldsymbol{I} + \frac{\boldsymbol{V} \boldsymbol{\Lambda_{\boldsymbol{Z}}}\Big[\boldsymbol{\Phi}^{\ast}\boldsymbol{T}, \boldsymbol{0}\Big]\boldsymbol{W}^{\ast}\boldsymbol{Q}\boldsymbol{W}^{}\Big[\boldsymbol{\Phi}^{\ast}\boldsymbol{T}, \boldsymbol{0}\Big]^{\ast}\boldsymbol{\Lambda}^{T}_{\boldsymbol{Z}}\boldsymbol{V}^{\ast}}{N_{0}}\bigg) \Big\} \label{eqn17} 
\end{eqnarray}
}
\vspace{-4mm}
\begin{eqnarray}
\text{s.t.} \quad \boldsymbol{Q} \ \succeq \ 0, \quad trace(\boldsymbol{Q}) \ \leq \ P_{0}. \label{eqn18}
\end{eqnarray}

\hspace{-5mm}
We perform the following sequence of substitutions in (\ref{eqn17}): 
$1) \ \boldsymbol{Q} \ = \ \boldsymbol{W}^{} \boldsymbol{Q}^{}_{1}\boldsymbol{W}^{\ast} \ \succeq \ \boldsymbol{0}$ \ and \ $\boldsymbol{Q}^{}_{1} \in \mathbb{C}^{N_{S} \times N_{S}}$, \\ 
$2) \ \boldsymbol{Q}^{}_{1} \ = \ \Big[\boldsymbol{Q}^{}_{2}, \ \boldsymbol{0}; \ \boldsymbol{0}, \ \boldsymbol{0}\Big] \ \succeq \ \boldsymbol{0}$ \ and \ $\boldsymbol{Q}^{}_{2} \in \mathbb{C}^{k \times k}$, \\ 
$3) \ \boldsymbol{Q}^{}_{2} \ = \ (\boldsymbol{\Phi}^{\ast}\boldsymbol{T})^{-1} \boldsymbol{Q}^{}_{3} \Big((\boldsymbol{\Phi}^{\ast}\boldsymbol{T})^{-1}\Big)^{\ast} \ \succeq \ \boldsymbol{0}$ \ and \ $\boldsymbol{Q}^{}_{3} \in \mathbb{C}^{k \times k}$. 

\hspace{-5mm}
With the above substitutions, (\ref{eqn17}) and (\ref{eqn18}) can be written 
in the following equivalent form:
\begin{eqnarray}
\max_{\boldsymbol{Q^{}_{}}, \ \boldsymbol{Q^{}_{1}},\ \boldsymbol{Q^{}_{2}}, \ \boldsymbol{Q^{}_{3}}} \bigg \{  \log_{2} \det\Big(\boldsymbol{I} + \frac{\boldsymbol{\Lambda_{\boldsymbol{H}}}\boldsymbol{Q}^{}_{3}\boldsymbol{\Lambda}^{T}_{\boldsymbol{H}}}{N_{0}}\Big)  \nonumber \\
- \log_{2} \det\Big(\boldsymbol{I} + \frac{\boldsymbol{\Lambda_{\boldsymbol{Z}}}\boldsymbol{Q}^{}_{3}\boldsymbol{\Lambda}^{T}_{\boldsymbol{Z}}}{N_{0}}\Big) \bigg \},  
\hspace{-4mm}
\label{eqn19} 
\end{eqnarray}
\begin{eqnarray}
\text{s.t.} \quad trace(\boldsymbol{Q}) \leq P_{0}, \
\boldsymbol{Q} \ = \ \boldsymbol{W}^{} \boldsymbol{Q}^{}_{1}\boldsymbol{W}^{\ast} , \nonumber \\
\boldsymbol{Q}^{}_{1} = \Big[\boldsymbol{Q}^{}_{2}, \boldsymbol{0}; \ \boldsymbol{0}, \boldsymbol{0}\Big] , \
\boldsymbol{Q}^{}_{2} = (\boldsymbol{\Phi}^{\ast}\boldsymbol{T})^{-1} \boldsymbol{Q}^{}_{3} \Big((\boldsymbol{\Phi}^{\ast}\boldsymbol{T})^{-1}\Big)^{\ast} , \nonumber \\
\boldsymbol{Q}^{}_{3} \succeq \boldsymbol{0}.
\label{eqn20}
\end{eqnarray}

Let there be $r$ non-zero diagonal entries in 
$\boldsymbol{\Lambda}_{\boldsymbol{H}}$. Since 
$\boldsymbol{\Lambda}_{\boldsymbol{H}}$ and $\boldsymbol{\Lambda}_{\boldsymbol{Z}}$
are diagonal matrices, (\ref{eqn19}) will be maximized if $\boldsymbol{Q}_{3}$ 
is selected to be of the following form:
\begin{eqnarray}
\boldsymbol{Q}^{}_{3} \ = \ \Big[\boldsymbol{Q}^{}_{4}, \ \boldsymbol{0}; \ \boldsymbol{0}, \ \boldsymbol{0}\Big] \ \succeq \boldsymbol{0},  \label{eqn21}
\end{eqnarray}
where $\boldsymbol{Q}_{4} \succeq \boldsymbol{0}$ and 
$\boldsymbol{Q}^{}_{4} \in \mathbb{C}^{r \times r}$.
In order to simplify the analysis further, we assume that 
$\boldsymbol{Q}_{4}$ is a diagonal matrix with 
$\boldsymbol{Q}_{4} = diag\big([q_1,q_2,\cdots,q_r]^{T}\big)$. 
Substituting (\ref{eqn21}) in (\ref{eqn19}) and (\ref{eqn20}), we
can write 
\begin{eqnarray}
\max_{\boldsymbol{Q^{}_{}}, \ \boldsymbol{Q^{}_{1}},\ \boldsymbol{Q^{}_{2}}, \ \boldsymbol{Q^{}_{3}}, \ \boldsymbol{Q^{}_{4}} \atop{q_1,q_2,\cdots,q_r}} \ \bigg \{ \ \log_{2} \det\Big(\boldsymbol{I} + \frac{\boldsymbol{\Lambda}^{r \times r}_{\boldsymbol{H}}\boldsymbol{Q}^{}_{4}\boldsymbol{\Lambda}^{^{r \times r} T}_{\boldsymbol{H}}}{N_{0}}\Big) - \nonumber \\ 
\log_{2} \det\Big(\boldsymbol{I} + \frac{\boldsymbol{\Lambda}^{r \times r}_{\boldsymbol{Z}}\boldsymbol{Q}^{}_{4}\boldsymbol{\Lambda}^{^{r \times r}T}_{\boldsymbol{Z}}}{N_{0}}\Big) \ \bigg \},  \label{eqn22} 
\end{eqnarray}
\begin{eqnarray}
\text{s.t.} \quad trace(\boldsymbol{Q}) \ \leq \ P_{0}, \quad
\boldsymbol{Q} = \boldsymbol{W}^{} \boldsymbol{Q}^{}_{1}\boldsymbol{W}^{\ast} , \nonumber \\
\boldsymbol{Q}^{}_{1} = \Big[\boldsymbol{Q}^{}_{2}, \boldsymbol{0}; \ \boldsymbol{0}, \boldsymbol{0}\Big] , \
\boldsymbol{Q}^{}_{2} = (\boldsymbol{\Phi}^{\ast}\boldsymbol{T})^{-1} \boldsymbol{Q}^{}_{3} \Big((\boldsymbol{\Phi}^{\ast}\boldsymbol{T})^{-1}\Big)^{\ast} , \nonumber \\
\boldsymbol{Q}^{}_{3} = \Big[\boldsymbol{Q}^{}_{4}, \boldsymbol{0}; \ \boldsymbol{0}, \boldsymbol{0}\Big], \ \boldsymbol{Q}^{}_{4} = diag\Big([q_1,\cdots,q_r]^{T}\Big) \succeq \boldsymbol{0}, 
\label{eqn23}
\end{eqnarray}
where $\boldsymbol{\Lambda}^{r \times r}_{\boldsymbol{H}} = diag\big([\lambda^{H}_{1},\lambda^{H}_{2},\cdots,\lambda^{H}_{r}]^{T}\big)$ and 
$\boldsymbol{\Lambda}^{r \times r}_{\boldsymbol{Z}} = diag\big([\lambda^{Z}_{1},\lambda^{Z}_{2},\cdots,\lambda^{Z}_{r}]^{T}\big)$ 
are leading $r \times r$ diagonal matrices of
$\boldsymbol{\Lambda}^{}_{\boldsymbol{H}}$ and $\boldsymbol{\Lambda}^{}_{\boldsymbol{Z}}$, respectively.

Rewrite the objective function in (\ref{eqn22}) in the following equivalent 
form:
\begin{eqnarray}
\max_{\boldsymbol{Q^{}_{}}, \ \boldsymbol{Q^{}_{1}},\ \boldsymbol{Q^{}_{2}}, \ \boldsymbol{Q^{}_{3}}, \ \boldsymbol{Q^{}_{4}} \atop{q_1,q_2,\cdots,q_r}} \ \sum^{r}_{i = 1} \bigg \{ \ \log_{2} \Big(1 + \frac{(\lambda^{H}_{i})^{2}q^{}_{i}}{N_{0}}\Big) - \nonumber \\ \log_{2}\Big(1 + \frac{(\lambda^{Z}_{i})^{2}q^{}_{i}}{N_{0}}\Big)\bigg \},  \label{eqn24} 
\end{eqnarray}
\hspace{20mm}
s.t. all constraints in (\ref{eqn23}).

\vspace{3mm}
We note that for $\lambda^{H}_{i} > \lambda^{Z}_{i}$, the function 
$\big \{ \log_{2} \big(1 + \frac{(\lambda^{H}_{i})^{2}q^{}_{i}}{N_{0}}\big) - \log_{2}\big(1 + \frac{(\lambda^{Z}_{i})^{2}q^{}_{i}}{N_{0}}\big) \big \}$ 
in (\ref{eqn24}) is positive, strictly increasing, and concave in the 
variable $q_{i} > 0$. Let $l \leq r$ be the number of $\lambda^{H}_{i}$s 
which are strictly greater than $\lambda^{Z}_{i}$s. We keep the $l$ terms 
in the summation in (\ref{eqn24}) for which $\lambda^{H}_{i} > \lambda^{Z}_{i}$ 
and remaining $r-l$ terms are discarded since they will not lead to positive 
secrecy rate. With this, the optimization problem (\ref{eqn24}) is written as 
follows:
\begin{eqnarray}
\max_{\boldsymbol{Q^{}_{}}, \ \boldsymbol{Q^{}_{1}},\ \boldsymbol{Q^{}_{2}}, \ \boldsymbol{Q^{}_{3}}, \ \boldsymbol{Q^{}_{4}} \atop{q_1,q_2,\cdots,q_l}} \ \sum^{l}_{i = 1} \bigg \{ \ \log_{2} \Big(1 + \frac{(\lambda^{H}_{i})^{2}q^{}_{i}}{N_{0}}\Big) - \nonumber \\ \log_{2}\Big(1 + \frac{(\lambda^{Z}_{i})^{2}q^{}_{i}}{N_{0}}\Big)\bigg \},  \label{eqn26} 
\end{eqnarray}
\begin{eqnarray}
\text{s.t.} \quad trace(\boldsymbol{Q}) \ \leq \ P_{0}, \quad
\boldsymbol{Q} = \boldsymbol{W}^{} \boldsymbol{Q}^{}_{1}\boldsymbol{W}^{\ast} , \nonumber \\
\boldsymbol{Q}^{}_{1} = [\boldsymbol{Q}^{}_{2}, \ \boldsymbol{0}; \ \boldsymbol{0}, \ \boldsymbol{0}] , \
\boldsymbol{Q}^{}_{2} = (\boldsymbol{\Phi}^{\ast}\boldsymbol{T})^{-1} \boldsymbol{Q}^{}_{3} \Big((\boldsymbol{\Phi}^{\ast}\boldsymbol{T})^{-1}\Big)^{\ast} , \nonumber \\
\boldsymbol{Q}^{}_{3} = [\boldsymbol{Q}^{}_{4}, \ \boldsymbol{0}; \ \boldsymbol{0}, \ \boldsymbol{0}], \nonumber \\
\boldsymbol{Q}^{}_{4} = diag([q_1,\cdots,q_l,0,\cdots,0]^{T}) \succeq \boldsymbol{0}. 
\label{eqn27}
\end{eqnarray}
The objective function in (\ref{eqn26}) is a sum of $l$ concave functions and 
all the constraints in (\ref{eqn27}) are linear. The above optimization problem 
is a concave maximization problem and it can be solved using nonlinear 
optimization techniques. We denote the optimum values of $q_1,q_2,\cdots,q_l$ 
obtained from (\ref{eqn26}) as $q^{g}_1,q^{g}_2,\cdots,q^{g}_l$, respectively.

$\bf{Remarks \ :}$
\begin{itemize}
\item 
A possible suboptimal approach to solve the optimization problem (\ref{eqn26}) 
will be to assign equal weights to all $q^{}_1,q^{}_2,\cdots,q^{}_l$, i.e.,
$q^{}_1=q^{}_2=\cdots=q^{}_l$, and solve the following optimization problem:
\begin{eqnarray}
\max_{\boldsymbol{Q^{}_{}}, \ \boldsymbol{Q^{}_{1}},\ \boldsymbol{Q^{}_{2}}, \ \boldsymbol{Q^{}_{3}}, \ \boldsymbol{Q^{}_{4}} \atop{q_1,q_2,\cdots,q_l}} \ trace(\boldsymbol{Q}) \nonumber \\
\text{s.t.} \quad trace(\boldsymbol{Q}) \ \leq \ P_{0}, \quad
\boldsymbol{Q} = \boldsymbol{W}^{} \boldsymbol{Q}^{}_{1}\boldsymbol{W}^{\ast} , \nonumber \\
\boldsymbol{Q}^{}_{1} = [\boldsymbol{Q}^{}_{2}, \ \boldsymbol{0}; \ \boldsymbol{0}, \ \boldsymbol{0}] , \
\boldsymbol{Q}^{}_{2} = (\boldsymbol{\Phi}^{\ast}\boldsymbol{T})^{-1} \boldsymbol{Q}^{}_{3} \Big((\boldsymbol{\Phi}^{\ast}\boldsymbol{T})^{-1}\Big)^{\ast} , \nonumber \\
\boldsymbol{Q}^{}_{3} = [\boldsymbol{Q}^{}_{4}, \ \boldsymbol{0}; \ \boldsymbol{0}, \ \boldsymbol{0}], \nonumber \\
\boldsymbol{Q}^{}_{4} = diag([q_1,\cdots,q_l,0,\cdots,0]^{T}) \succeq \boldsymbol{0}, \nonumber \\
q^{}_1=q^{}_2=\cdots=q^{}_l. \nonumber
\end{eqnarray}
\item 
We note that the MIMO wiretap problem in (\ref{eqn26}) with the total available 
transmit power constraint, $trace(\boldsymbol{Q}) \ \leq \ P_{0}$, in $(\ref{eqn27})$ 
can also be extended to the scenario when there is an individual power constraint on 
$\boldsymbol{Q}$, i.e., $diag(\boldsymbol{Q}) \leq [P_1,P_2,\cdots,P_{N_S}]^{T}$, where 
$P_1,P_2,\cdots,P_{N_S}$ are the available transmit powers for antennas $1,2,\cdots,N_{s}$, 
respectively.
\end{itemize}
\section{MIMO Wiretap Channel with Finite-Alphabet Input}
\label{sec4}
The optimization problem (\ref{eqn26}) can be equivalently viewed as the sum 
secrecy rate of $l$ scalar Gaussian wiretap channels with power constraints 
in (\ref{eqn27}). $\sqrt{\frac{(\lambda^{H}_{i})^{2}q_i}{N_0}}$ and 
$\sqrt{\frac{(\lambda^{Z}_{i})^{2}q_i}{N_0}}$ correspond to the destination
and eavesdropper channel coefficients, respectively, associated with the 
$i$th Gaussian wiretap channel where $1 \leq i \leq l$ and noise 
$\sim \mathcal{CN}(0, 1)$. In this section, we consider the power 
allocation scheme for the above channel model when the input to each 
scalar wiretap channel is from a finite alphabet set 
$\mathbb{A} = \{a_1, a_2,\cdots,a_M\}$ of size $M$. We assume that 
symbols from the set $\mathbb{A}$ are drawn equiprobably and 
$\mathbb{E} \{ \lvert a \rvert^{2}\} = 1$. With finite-alphabet 
input, we write the optimization problem (\ref{eqn26}) as follows:
\begin{eqnarray}
\max_{\boldsymbol{Q^{}_{}}, \ \boldsymbol{Q^{}_{1}},\ \boldsymbol{Q^{}_{2}}, \ \boldsymbol{Q^{}_{3}}, \ \boldsymbol{Q^{}_{4}} \atop{q_1,q_2,\cdots,q_l}} \ \sum^{l}_{i = 1} \bigg \{I\Big({\frac{(\lambda^{H}_{i})^{2}q_i}{N_0}}\Big) - I\Big({\frac{(\lambda^{Z}_{i})^{2}q_i}{N_0}}\Big) \bigg \},  \label{eqn28} 
\end{eqnarray}
\hspace{2cm} s.t. all constraints in (\ref{eqn27}). 

\hspace{-5mm}
$I(.)$ in (\ref{eqn28}) is the mutual information function with finite-alphabet 
input and it is explicitly written as follows:
\begin{eqnarray}
I(\rho) & = & \frac{1}{M} \sum\limits^{M}_{i = 1} \int p_{n}\big(z_{_{}} - \sqrt{\rho} a_{i}\big)  \nonumber \\
& & . \log_{2} \frac{p_{n}(z_{_{}} - \sqrt{\rho} a_{i})}{\frac{1}{M} \sum\limits^{M}_{m = 1}p_{n}(z_{_{}} - \sqrt{\rho} a_{m})} d z_{_{}}, \label{eqn30}
\end{eqnarray}
where $p_n(\theta) = \frac{1}{\pi} e^{{{-\mid \theta \mid}^{2}}}$. 
Solving the optimization problem (\ref{eqn28}) for optimum $q_1,q_2,\cdots,q_l$ 
is hard. A suboptimal approach to find the secrecy rate with finite-alphabet 
input will be to use $q^{g}_1,q^{g}_2,\cdots,q^{g}_l$ directly in (\ref{eqn28}) 
obtained from (\ref{eqn26}) with Gaussian input. This suboptimal approach to 
find the secrecy rate with finite-alphabet input could be adverse 
and it could lead to reduced secrecy rate without transmit power control.
In the Appendix, we show that the secrecy rate with finite-alphabet input 
for a Gaussian wiretap channel is a unimodal function in transmit power, 
i.e., there exist a unique transmit power at which the secrecy rate attains 
its maximum value. 

Let $q^{ul}_1,q^{ul}_2,\cdots,q^{ul}_l$ be the upper limit for $q_1,q_2,\cdots,q_l$ 
obtained using the method proposed in the Appendix. Using these upper limits 
$q^{ul}_1,q^{ul}_2,\cdots,q^{ul}_l$, we rewrite the optimization problem 
(\ref{eqn26}) as follows:
\begin{eqnarray}
\max_{\boldsymbol{Q^{}_{}}, \ \boldsymbol{Q^{}_{1}},\ \boldsymbol{Q^{}_{2}}, \ \boldsymbol{Q^{}_{3}}, \ \boldsymbol{Q^{}_{4}} \atop{q_1,q_2,\cdots,q_l}} \ \sum^{l}_{i = 1} \bigg \{ \ \log_{2} \Big(1 + \frac{(\lambda^{H}_{i})^{2}q^{}_{i}}{N_{0}}\Big) - \nonumber \\ \log_{2}\Big(1 + \frac{(\lambda^{Z}_{i})^{2}q^{}_{i}}{N_{0}}\Big)\bigg \},  \label{eqn31} 
\end{eqnarray}
\begin{eqnarray}
\text{s.t.} \quad trace(\boldsymbol{Q}) \ \leq \ P_{0}, \ \boldsymbol{Q} \ = \ \boldsymbol{W}^{} \boldsymbol{Q}^{}_{1}\boldsymbol{W}^{\ast} , \nonumber \\
\boldsymbol{Q}^{}_{1} = [\boldsymbol{Q}^{}_{2}, \boldsymbol{0}; \ \boldsymbol{0}, \boldsymbol{0}] , \ \boldsymbol{Q}^{}_{2} = (\boldsymbol{\Phi}^{\ast}\boldsymbol{T})^{-1} \boldsymbol{Q}^{k \times k}_{3} ((\boldsymbol{\Phi}^{\ast}\boldsymbol{T})^{-1})^{\ast} , \nonumber \\
\boldsymbol{Q}^{}_{3} \ = \ [\boldsymbol{Q}^{}_{4}, \ \boldsymbol{0}; \ \boldsymbol{0}, \ \boldsymbol{0}] , \nonumber \\
\boldsymbol{Q}^{}_{4} = diag([q_1,q_2,\cdots,q_l,0,\cdots,0]^{T}) \ \succeq \boldsymbol{0}, \nonumber \\
{[q_1,q_2,\cdots,q_l]}^{T} \leq [q^{ul}_1,q^{ul}_2,\cdots,q^{ul}_l]^T.
\label{eqn32}
\end{eqnarray}
We denote the optimum solution of (\ref{eqn31}) as $q^{f}_1,q^{f}_2,\cdots,q^{f}_l$. 
If $q^{f}_1,q^{f}_2,\cdots,q^{f}_l$ are used in (\ref{eqn28}) to compute the 
secrecy rate with finite-alphabet input, it will not lead to reduced secrecy 
rate due to the presence of additional constraint 
$[q_1,q_2,\cdots,q_l]^{T} \leq [q^{ul}_1,q^{ul}_2,\cdots,q^{ul}_l]^T$ in 
(\ref{eqn32}). We will see this in the numerical results presented in the
next section.

\section{Results and Discussions}
\label{sec5}
\begin{figure}
\hspace{-2mm}
\includegraphics[totalheight=3.0in,width=3.75in]{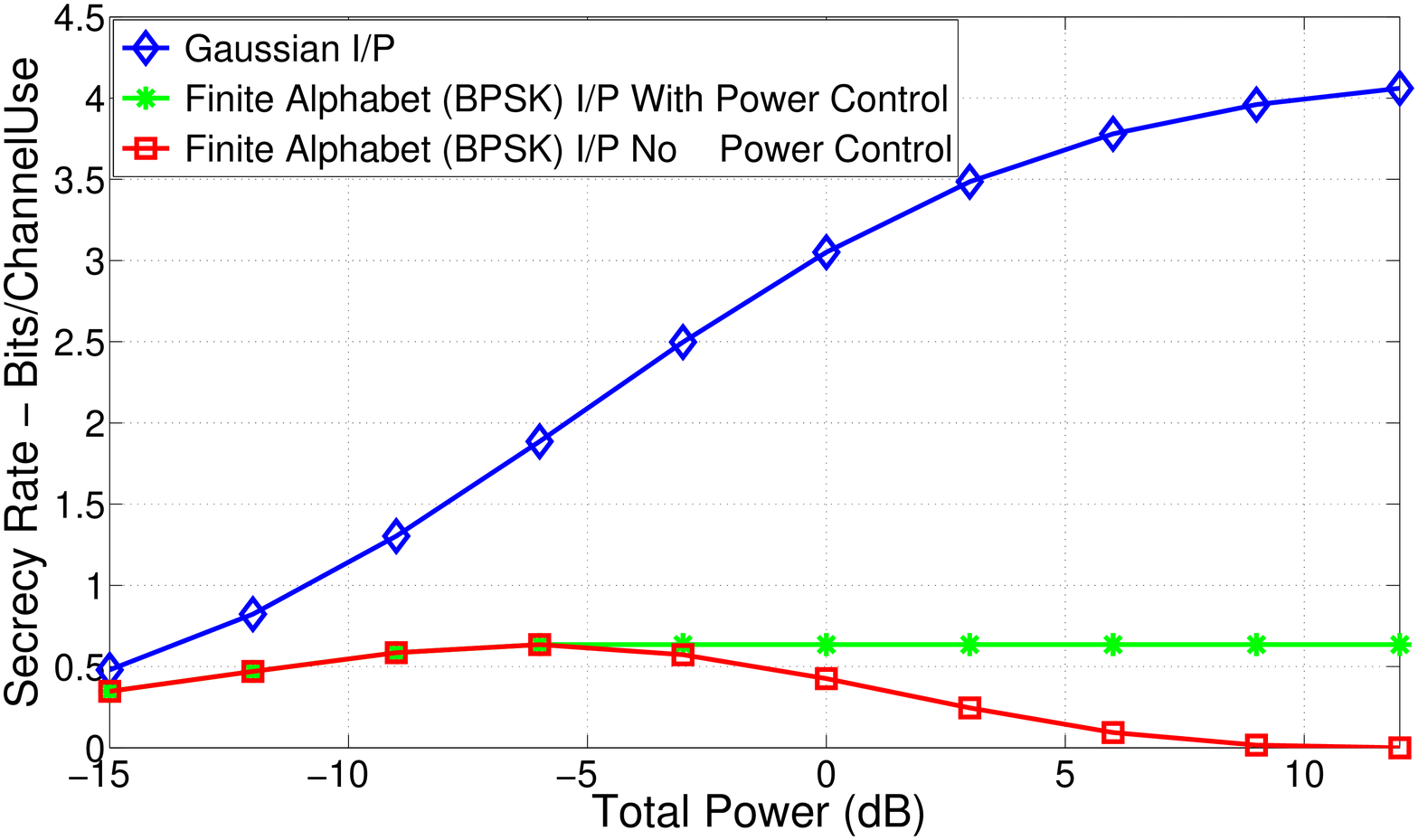}
\vspace{-6mm}
\caption{Secrecy rate vs total power of MIMO wiretap channel with known 
destination CSI and unknown (statistical) eavesdroppers CSI. 
$N_{S}=N_{D}=N_{E_{j}}=3, N_{0}=1, \sigma^{}_{E_{j_{0}}}=0.5$. }
\vspace{-2mm}
\label{fig2}
\end{figure}

We computed the secrecy rate for MIMO wiretap channel with $N_{S}=N_{D}=N_{E_{j}}=3$ 
(i.e., source, destination and eavesdroppers have 3 antennas each) by simulations.
We take that $N_{0} = 1$, $\sigma^{}_{E_{j_{0}}} = 0.5$, and 
\begin{eqnarray}
\boldsymbol{H} = \left[\footnotesize \begin{array}{cc}0.0799 - 0.1191i, \  1.9709 + 0.2753i, \ -0.8066 + 0.8648i \\ 0.3111 - 0.1545i, \ -0.8250 + 0.5312i, \ -0.7731 - 0.9074i \\ 0.0719 + 0.3828i, \ -1.3112 + 1.2574i, \ -0.3066 - 1.6468i
\end{array} \right]. \nonumber
\end{eqnarray}
We computed the secrecy rate for three different cases: 
\begin{itemize}
\item
{\em Case 1:} The secrecy rate is computed with Gaussian input. 
\item
{\em Case 2:} The secrecy rate is computed with binary alphabet (BPSK) input but 
with no power control, i.e., the solution obtained directly from (\ref{eqn26}) 
is used to compute the finite-alphabet secrecy rate in (\ref{eqn28}). 
\item
{\em Case 3:} The secrecy rate is computed with binary alphabet (BPSK) input but 
with power control, i.e., the solution obtained from (\ref{eqn31}) is used to 
compute the finite-alphabet secrecy rate in (\ref{eqn28}).
\end{itemize}
The computed secrecy rate results for the above three cases are shown in 
Fig. \ref{fig2}. From Fig. \ref{fig2}, it can be seen that, as expected, 
the secrecy rate for MIMO wiretap channel with Gaussian alphabet input 
({\em Case 1}) increases with increase in $P_{0}$. The secrecy rate 
with BPSK input but with no power control ({\em Case 2}) first increases 
with increase in $P_{0}$ and then decreases to zero at high transmit powers. 
This is due to the fact that at high transmit powers with finite-alphabet input, 
the information rate at the eavesdroppers equals the information rate at the 
destination which causes the secrecy rate go to zero. However, when the
power allocation scheme proposed in Section \ref{sec4} is used, the 
MIMO wiretap secrecy rate with BPSK input ({\em Case 3}) does not go to
zero at high transmit powers (as was observed in Case 2). Instead, the 
secrecy rate increases with increasing transmit power and remains flat at 
some non-zero secrecy rate at high transmit powers. This is because of the 
presence of the additional power constraint
$[q_1,q_2,\cdots,q_l]^{T} \leq [q^{ul}_1,q^{ul}_2,\cdots,q^{ul}_l]^T$, in 
(\ref{eqn32}). 

\section{Conclusions}
\label{sec6}
We studied the problem of power allocation for secrecy in MIMO wiretap channel
with finite-alphabet input. Our work differed from past works in the following 
aspects: we assumed that only the statistical knowledge of the eavesdropper CSI 
is known, and we considered multiple eavesdroppers. To study the problem, we
first considered the MIMO wiretap channel with Gaussian input, where we 
transformed the secrecy rate max-min optimization problem to a concave 
maximization problem which maximized the sum secrecy rate of $l$ scalar 
wiretap channels subject to linear constraints on the transmit covariance 
matrix. When the transmit covariance matrix obtained in the Gaussian
input setting is used in the finite-alphabet input setting, the secrecy 
rate decreased for increasing transmit powers leading to zero secrecy
rate at high transmit powers. To alleviate this secrecy rate loss, we
proposed a power allocation scheme using an additional power
constraint in the problem. The proposed power allocation scheme was shown to
alleviate the secrecy rate loss problem and achieve flat non-zero secrecy
rate at high transmit powers.

\appendix
In this appendix, we show that the secrecy rate with finite-alphabet input 
for a Gaussian wiretap channel is a unimodal function in transmit power, 
i.e., there exist a unique transmit power at which secrecy rate attains 
its maximum value. Let $y_{D}$ and $y_{E}$ be the received signals at 
the destination and eavesdropper, respectively, in a Gaussian wiretap 
channel, i.e.,
\begin{eqnarray}
y_{D} = \sqrt{P} hx + \eta_{D} \label{eqn33} \\
y_{E} = \sqrt{P} zx + \eta_{E}, \label{eqn34}
\end{eqnarray}
where $h$ and $z$ are known channel coefficients for the destination and 
eavesdropper channels, respectively, $x$ is the transmitted source symbol 
from a finite-alphabet set $\mathbb{A}$ as described in Section \ref{sec4} 
with $\mathbb{E} \{ \lvert x \rvert^{2} \} = 1$, $P$ is the power transmitted 
by the source, and $\eta_{D}$ and $\eta_{E}$ are the independent additive 
noise terms at the destination and eavesdropper $\sim \mathcal{CN}(0, 1)$.

Using (\ref{eqn33}), the information rate at the destination, $R^{f}_{D}$, 
with finite-alphabet input is 
\begin{eqnarray}
R^{f}_{D} = I(\lvert h \rvert^{2} P). 
\label{eqn41}
\end{eqnarray}
Similarly, using (\ref{eqn34}), the information rate at the eavesdropper, 
$R^{f}_{E}$, with finite-alphabet input is
\begin{eqnarray}
R^{f}_{E} = I(\lvert z \rvert^{2} P). 
\label{eqn42}
\end{eqnarray}
$I(.)$ in (\ref{eqn41}) and (\ref{eqn42}) is the mutual information function 
as defined in (\ref{eqn30}). The secrecy rate, $R^{f}_{s}$, with finite-alphabet 
input for the Gaussian wiretap channel is obtained as
\begin{eqnarray}
R^{f}_{s}=R^{f}_{D}-R^{f}_{E}=I(\lvert h \rvert^{2} P)-I(\lvert z \rvert^{2} P). 
\label{eqn35}
\end{eqnarray}
With $P > 0$, $R^{f}_{s}$ in (\ref{eqn35}) will be positive only when 
$\lvert h \rvert > \lvert z \rvert$. Therefore, w.l.o.g. we assume that 
$\lvert h \rvert > \lvert z \rvert$. Using Theorem 1 in \cite{ic15} to find 
the derivatives of $R^{f}_{D}$ and $R^{f}_{E}$ w. r. t. $P$, respectively, 
we get
\begin{eqnarray}
{\frac{ dR^{f}_{D}}{dP_{}} = \lvert h \rvert^{2} \mbox{\scriptsize{MMSE}}\big(\lvert h \rvert^{2} P\big) \log_{2}{e}}   \label{eqn43}
\end{eqnarray}
\begin{eqnarray}
{\frac{ dR^{f}_{E}}{dP_{}} = \lvert z \rvert^{2} \mbox{\scriptsize{MMSE}}\big(\lvert z \rvert^{2} P\big) \log_{2}{e}}.   \label{eqn44}
\end{eqnarray}
Using (\ref{eqn43}) and (\ref{eqn44}), taking the derivative of $R^{f}_{s}$ w.r.t. 
$P$ and equating it to zero, we get
\begin{eqnarray}
{\frac{ dR^{f}_{s}}{dP_{}} = \Big( \lvert h \rvert^{2} \mbox{\scriptsize{MMSE}}\big(\lvert h \rvert^{2} P\big)-\lvert z \rvert^{2} \mbox{\scriptsize{MMSE}}\big(\lvert z \rvert^{2} P\big) \Big ) \log_{2}{e}} \nonumber \\
= 0.  \label{eqn36}
\end{eqnarray}
We intend to seek the solution, $P = P_{opt}$, of (\ref{eqn36}). We show that, 
with finite alphabet, this solution is unique and secrecy rate, $R^{f}_{s}$, 
attains it's maximum value at $P = P_{opt}$.

\begin{figure}
\hspace{-4mm}
\includegraphics[totalheight=3.05in,width=3.75in]{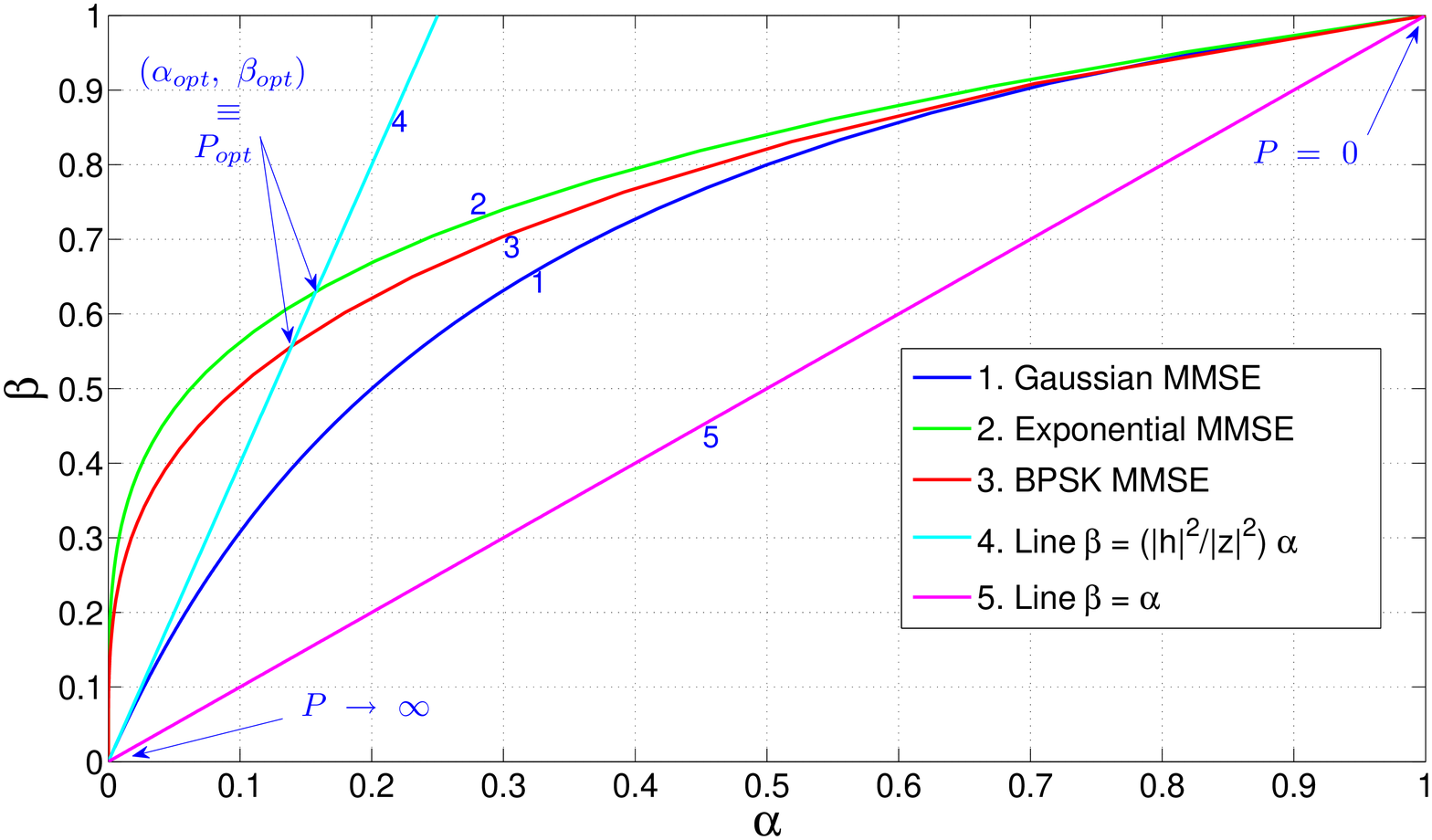}
\vspace{-6mm}
\caption{Various MMSE $\beta$ vs $\alpha$ curves with $\lvert h \rvert^{2} = 2.0$ 
and $\lvert z \rvert^{2} = 0.5$.}
\vspace{-2mm}
\label{fig3}
\end{figure}

For various $M$-ary alphabets, it is shown in \cite{ic15, ic16} that 
$1)$ $\mbox{\small{MMSE}}$ is a positive, strictly monotonic decreasing 
function in $\mbox{\small{SNR}}$ and in the limit approaches zero as  
$\mbox{\small{SNR}}$ tends to infinity, and $2)$ at high $\mbox{\small{SNR}}$s, 
$\mbox{\small{MMSE}}$ decreases exponentially (Theorems 3 and 4 in \cite{ic15}). 
Since $\mbox{\small{MMSE}}$ is a strictly monotonic decreasing function, 
it's inverse, ${\mbox{\small{MMSE}}}^{-1}$, exists. Define
\begin{eqnarray}
\alpha \ = \ \mbox{\footnotesize{MMSE}}(\lvert h \rvert^{2} P) \ \Longrightarrow \ P \ = \ \frac{1}{\lvert h \rvert^{2}} {\mbox{\footnotesize{MMSE}}}^{-1}(\alpha), 
\label{eqn37} \\
\text{and} \quad
\beta \ = \ \mbox{\footnotesize{MMSE}}(\lvert z \rvert^{2} P).  
\label{eqn38}
\end{eqnarray}
Using (\ref{eqn37}), we rewrite  (\ref{eqn38}) in terms of $\alpha$ as
\begin{eqnarray}
\beta \ = \ \mbox{\footnotesize{MMSE}}(\lvert z \rvert^{2} P) \ =  \ \mbox{\footnotesize{MMSE}}\Big( \frac{\lvert z \rvert^{2}}{\lvert h \rvert^{2}} {\mbox{\footnotesize{MMSE}}}^{-1}(\alpha)\Big).
\label{eqn39}
\end{eqnarray}
It can be easily shown that $\beta$ is a strictly monotonic increasing function 
in $\alpha$. We plot $\beta$ as a function of $\alpha$ for three different 
$\mbox{\footnotesize{MMSE}}$ functions and two straight lines in Fig. \ref{fig3}.
Point $(\alpha,\beta)=(0,0) \equiv O$ in the plot corresponds to 
$P_{} \rightarrow \infty$. Similarly, point $(\alpha,\beta)=(1,1)$ corresponds 
to $P_{} = 0$. 

\vspace{2mm}
$1) \textbf{\ Gaussian MMSE Function:}$ We take 
$\mbox{\footnotesize{MMSE}}(\lvert h \rvert^{2} P) = \frac{1}{(1 + \lvert h \rvert^{2} P)} = \alpha$ 
and 
$\mbox{\footnotesize{MMSE}}(\lvert z \rvert^{2} P) = \frac{1}{(1 + \lvert z \rvert^{2} P)} = \beta$. 
With this choice of $\mbox{\footnotesize{MMSE}}$ functions, 
$\beta = \frac{1}{\left(1 + \frac{\lvert z \rvert^{2}}{\lvert h \rvert^{2}} ({\frac{1}{\alpha} \ - \ 1})\right)}$. 
The slope of this curve at the origin, $(0,0)$, is 
\begin{eqnarray}
\frac{d \beta}{d \alpha} \ \text{at} \ (\alpha = 0) \ = \ \frac{\frac{d \beta}{d P_{}}}{\frac{d \alpha}{d P_{}}} \ \text{as} \ (P_{} \rightarrow \infty) \ = \ \frac{\lvert h \rvert^{2}}{\lvert z \rvert^{2}}. \nonumber 
\end{eqnarray}
This implies that 
$\beta=\big(\frac{\lvert h \rvert^{2}}{\lvert z \rvert^{2}}\big) \alpha$ 
is tangent to the Gaussian 
$\mbox{\footnotesize{MMSE}}$ $\beta \ \text{vs} \ \alpha$ curve at the origin 
$(0,0)$. 

\vspace{2mm}
$2) \textbf{\ Exponential MMSE Function:}$ 
We take $\mbox{\footnotesize{MMSE}}(\lvert h \rvert^{2} P) = \exp^{-(\lvert h \rvert^{2} P)} = \alpha$, 
and 
$\mbox{\footnotesize{MMSE}}(\lvert z \rvert^{2} P) = \exp^{-(\lvert z \rvert^{2} P)} = \beta$. 
With this choice of $\mbox{\footnotesize{MMSE}}$ functions, 
$\beta = {\alpha}^{(\frac{\lvert z \rvert^{2}}{\lvert h \rvert^{2}})}$. 
The $\beta$ axis, i.e., $\alpha = 0$, is tangent to the exponential 
$\mbox{\footnotesize{MMSE}}$ $\beta \ \text{vs} \ \alpha$ curve at the 
origin $(0,0)$. 

\vspace{2mm}
$3) \textbf{\ $M$-ary MMSE Functions:}$ At high $\mbox{\footnotesize{SNR}}$s, 
$\mbox{\footnotesize{MMSE}}$ for $M$-ary alphabets decreases exponentially 
(Theorems 3 and 4 in \cite{ic15}). This implies that the $\beta$ axis, i.e., 
$\alpha = 0$, is tangent to the $M$-ary 
$\mbox{\footnotesize{MMSE}}$ $\beta \ \text{vs} \ \alpha$ 
curve at the origin $(0, 0)$. 

\vspace{2mm}
$4) \textbf{\ Straight Line:}$ $\beta = \big(\frac{\lvert h \rvert^{2}}{\lvert z \rvert^{2}}\big) \alpha$. 

\vspace{2mm}
$5) \textbf{\ Straight Line:}$ $\beta = \alpha$. 

\vspace{2mm}
Since the $\beta$ axis, i.e., $\alpha = 0$, is a tangent to exponential 
$\mbox{\footnotesize{MMSE}}$ $\beta \ \text{vs} \ \alpha$ curve at the origin 
$(0, 0)$, the exponential $\mbox{\scriptsize{MMSE}}$ $\beta \ \text{vs} \ \alpha$ 
curve will always intersect with the
$\beta \ = \ \big(\frac{\lvert h \rvert^{2}}{\lvert z \rvert^{2}}\big) \alpha$ 
line at a point other than $(0, 0)$. This implies that for exponential
$\mbox{\footnotesize{MMSE}}$ function, there exists a $P = P_{opt}$ which makes 
(\ref{eqn36}) zero. Uniqueness of $P_{opt}$ can be confirmed by substituting 
exponential $\mbox{\footnotesize{MMSE}}$ function directly in (\ref{eqn36}).
Also, since $\lvert h \rvert^{} > \lvert z \rvert^{} \geq 0$, $R^{f}_{s}$ 
will attain it's maximum value at $P = P_{opt}$.

When the $\mbox{\footnotesize{MMSE}}$ function is Gaussian, the
Gaussian $\mbox{\footnotesize{MMSE}}$ $\beta \ \text{vs} \ \alpha$ curve, 
$\beta=\frac{1}{\left(1+\frac{\lvert z \rvert^{2}}{\lvert h \rvert^{2}} ({\frac{1}{\alpha}-1})\right)}$,  
does not intersect with $\beta \ = \ \big(\frac{\lvert h \rvert^{2}}{\lvert z \rvert^{2}}\big) \alpha$ 
line at any other point other than $(0, 0)$. In fact, the
$\beta = \big(\frac{\lvert h \rvert^{2}}{\lvert z \rvert^{2}}\big) \alpha$ 
line is tangent to the Gaussian 
$\mbox{\footnotesize{MMSE}}$ $\beta \ \text{vs} \ \alpha$ curve at $(0, 0)$.
This implies that for Gaussian $\mbox{\footnotesize{MMSE}}$, there is no 
$P = P_{opt}$ which makes (\ref{eqn36}) zero. This fact can also be confirmed 
by substituting the Gaussian $\mbox{\footnotesize{MMSE}}$ function directly 
in (\ref{eqn36}).

\begin{figure}
\center
\includegraphics[totalheight=3.05in,width=3.70in]{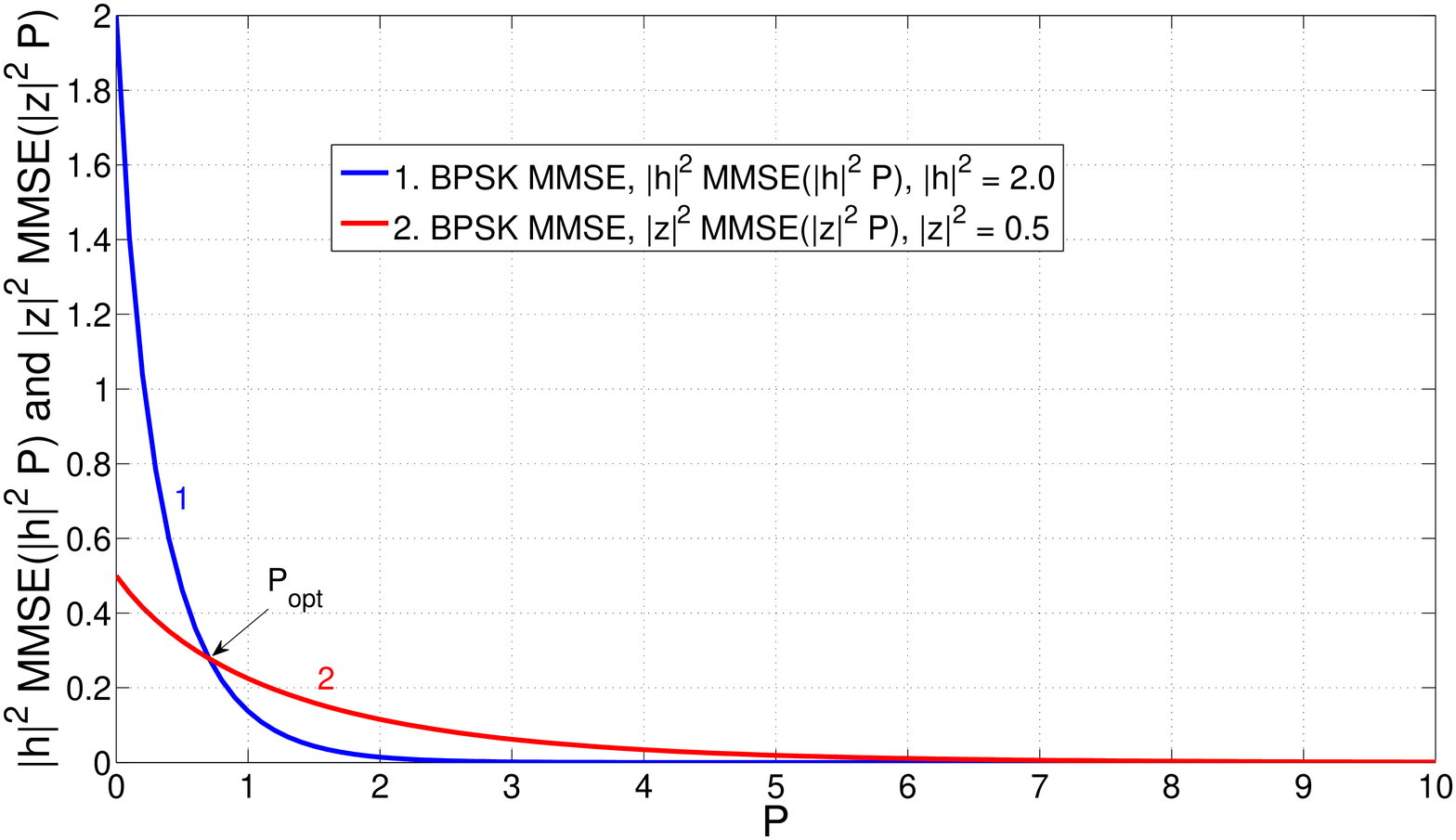}
\vspace{-6mm}
\caption{BPSK MMSE vs P curves with $\lvert h \rvert^{2} = 2.0$ and 
$\lvert z \rvert^{2} = 0.5$.}
\vspace{-4mm}
\label{fig4}
\end{figure}

The $\mbox{\footnotesize{MMSE}}$ function of $M$-ary alphabets at high 
$\mbox{\footnotesize{SNR}}$s decreases exponentially, which means $\beta$ axis, 
i.e., $\alpha=0$, is tangent to $M$-ary $\mbox{\footnotesize{MMSE}}$ 
$\beta \ \text{vs} \ \alpha$ curve at the origin $(0, 0)$. This implies that 
$M$-ary $\mbox{\footnotesize{MMSE}}$ $\beta \ \text{vs} \ \alpha$ curve will 
always intersect with 
$\beta = \big(\frac{\lvert h \rvert^{2}}{\lvert z \rvert^{2}}\big) \alpha$ 
line at a point other than $(0, 0)$. This shows that for $M$-ary 
$\mbox{\footnotesize{MMSE}}$ function, there exists a $P = P_{opt}$ which 
makes (\ref{eqn36}) zero. To prove the uniqueness of $P_{opt}$, let 
$\lvert h \rvert^{2} \mbox{\scriptsize{MMSE}}\big(\lvert h \rvert^{2} P\big)$ and
$\lvert z \rvert^{2} \mbox{\scriptsize{MMSE}}\big(\lvert z \rvert^{2} P\big)$ in 
(\ref{eqn36}) intersect for the first time at $P = P_{opt}$ from $P = 0$. Since 
$\lvert h \rvert > \lvert z \rvert \geq 0$, this implies that 
$\lvert h \rvert^{2} \mbox{\scriptsize{MMSE}}\big(\lvert h \rvert^{2} P\big) > \lvert z \rvert^{2} \mbox{\scriptsize{MMSE}}\big(\lvert z \rvert^{2} P\big)$
for all $P < P_{opt}$ and 
$\lvert h \rvert^{2} \mbox{\scriptsize{MMSE}}\big(\lvert h \rvert^{2} P\big) < \lvert z \rvert^{2} \mbox{\scriptsize{MMSE}}\big(\lvert z \rvert^{2} P\big)$
in some neighborhood of $P > P_{opt}$. Monotonicity of $\mbox{\small{MMSE}}$ 
\cite{ic16} implies that 
$\lvert h \rvert^{2} \mbox{\scriptsize{MMSE}}\big(\lvert h \rvert^{2} P\big)$ and
$\lvert z \rvert^{2} \mbox{\scriptsize{MMSE}}\big(\lvert z \rvert^{2} P\big)$ 
will not intersect for any finite $P > P_{opt}$. This can be seen in Fig. 
\ref{fig4} also. This proves the uniqueness of $P_{opt}$. The above analysis 
also implies that at $P = P_{opt}$ the secrecy rate $R^{s}_{f}$ will attain its 
maximum value.

\subsection{Numerical computation of $P_{opt}$}
We can find $P_{opt}$ of (\ref{eqn36}) for $M$-ary $\mbox{\footnotesize{MMSE}}$ 
functions using gradient based method as follows. 

\vspace{2mm}
$\bf{Step\ 1:}$ Let $P_{opt}$ lie in the interval 
$[P_{ll}, \ P_{ul}]$, $P_{ll} \geq 0$, $P_{ul} \leq P_{}$. Let $\epsilon$ be 
a small positive number. 

\vspace{2mm}
$\bf{Step \ 2:}$ $P_{opt} = (P_{ll} + P_{ul})/2$. Compute 
$\frac{ dR^{f}_{s}}{dP_{}}$ using (\ref{eqn36}) at $P_{opt}$. 

\vspace{2mm}
$\bf{Step \ 3:}$ If $\frac{ dR^{f}_{s}}{dP_{}} \geq 0$, then $P_{ll} = P_{opt}$; 
else $P_{ul} = P_{opt}$. 

\vspace{2mm}
Repeat $\bf{Step \ 2}$ and $\bf{Step \ 3}$ until $P_{ul} - P_{ll} \leq \delta$, 
where $\delta$ is a small positive number.


\begin{thebibliography}{99}
%
%
%
%
%
%
%
%
%
%
%
\vspace{0mm}
\bibitem{ic1}
S. Shafiee and S. Ulukus, ``Achievable rates in Gaussian MISO channels with
secrecy constraint,'' {\em Proc. IEEE ISIT'2007}, June 2007.


\vspace{0mm}
\bibitem{ic2}
A. Khisti and G. Wornell, ``Secure transmission with multiple antennas-I: The
MISOME wiretap channel,'' {\em IEEE Trans. Inf. Theory}, vol. 56, no. 7,
pp. 3088-3104, July 2010.

\vspace{0mm}
\bibitem{ic3}
F. Oggier and B. Hassibi, ``The secrecy capacity of the MIMO wiretap channel,'' 
{\em Proc. IEEE ISIT'2008}, July 2008.

\vspace{0mm}
\bibitem{ic4}
A. Khisti and G. Wornell, ``Secure transmission with multiple antennas-II: The
MIMOME wiretap channel,'' {\em IEEE Trans. Inf. Theory}, vol. 56, no. 7,
pp. 3088-3104, July 2010.

\vspace{0mm}
\bibitem{ic5}
J. Li and A. P. Petropulu, ``On ergodic secrecy rate for Gaussian MISO wiretap 
channels,'' {\em IEEE Trans. Wireless Commun.}, vol. 10, no. 4, pp. 1176-1187, 
April 2011.

\vspace{0mm}
\bibitem{ic6}
J. Liu, Y. T. Hou, and H. D. Sherali ``Optimal power allocation for achieving 
perfect secrecy capacity in MIMO wire-tap channels,'' {\em Proc. CISS'2009}, 
March 2009.

\vspace{0mm}
\bibitem{ic7}
S. A. A. Fakoorian and A. L. Swindlehurst, ``Full rank solutions for the MIMO 
Gaussian wiretap channel with an average power constraint,'' {\em IEEE Trans. 
Signal Process.}, vol. 61, no. 10, pp. 2620-2631, May 2013.

\bibitem{ic8}
M. R. D. Rodrigues, A. S. Baruch, and M. Bloch, ``On Gaussian wiretap channels 
with M-PAM inputs,'' {\em 2010 European Wireless Conference}, pp.774-781, 
April 2010.

\bibitem{ic9}
G. D. Raghava and B. S. Rajan, ``Secrecy capacity of the Gaussian wire-tap 
channel with finite complex constellation input,'' Online: 
arXiv:1010.1163v1 [cs.IT] 6 Oct 2010.

\bibitem{ic10}
S. Bashar, Z. Ding, and C. Xiao, ``On the secrecy rate of multi-antenna 
wiretap channel under finite-alphabet input,'' {\em IEEE Commun. Letters}, 
vol. 15, no. 5, pp. 527-529, May 2011.

\bibitem{ic11}
S. Vishwakarma and A. Chockalingam, ``Decode-and-forward relay beamforming for 
secrecy with finite-alphabet input,'' {\em IEEE Commun. Letters}, vol. 17, 
no. 5, pp. 912-915, May 2013.

\bibitem{ic12}
Y. Wu, C. Xiao, Z. Ding, X. Gao, and S. Jin, ``Linear precoding for finite-alphabet 
signaling over MIMOME wiretap channels,'' {\em IEEE Trans. Veh. Tech.}, vol. 61, 
no. 6, pp. 2599-2612, July 2012.

%
%
%
%
%
%
%

\vspace{0mm}
\bibitem{ic13}
C. Paige and M. A. Saunders, ``Towards a generalized singular value decomposition,'' 
{\em SIAM J. Numer. Anal.}, vol. 18, no. 3, pp. 398-405, June 1981.


\vspace{0mm}
\bibitem{ic15}
A. Lozano, A. M. Tulino, and S. Verdu, ``Optimum power allocation for parallel 
Gaussian channels with arbitrary input distributions,'' {\em IEEE Trans. Inf. 
Theory}, vol. 52, no. 7, pp. 3033-3051, July 2006.

\vspace{0mm}
\bibitem{ic16}
D. Guo, Y. Wu, S. Shamai, and S. Verdu, ``Estimation in Gaussian noise: properties 
of the minimum mean-square error,'' {\em IEEE Trans. Inf. Theory}, vol. 57, no. 4, 
pp. 2371-2385, April 2011.

\vspace{0mm}
\bibitem{ic17}
S. Boyd and L. Vandenberghe, {\em Convex optimization}, Cambridge Univ. Press, 2004.



\end{thebibliography}
\end{document}